\documentclass[a4paper]{spie}  
\usepackage{amssymb,amsmath,psfrag,url}
\usepackage[config,footnotesize,bf]{caption}
\usepackage{graphicx}
\usepackage[config]{subfig}
\usepackage[mediumspace,mediumqspace,squaren]{SIunits}
\usepackage{psfrag}
\usepackage{cite}

\newcommand{\Field}[1]{{{#1}}}
\newcommand{\Tensor}[1]{{{#1}}}
\newcommand{\curl}{\nabla\times}

\title{3D finite element simulation of optical modes in VCSELs}

\author{
Maria Rozova,\supit{\,a}
Jan Pomplun,\supit{\,b}
Lin Zschiedrich,\supit{\,b}
Frank Schmidt,\supit{\,ab}
Sven Burger\supit{\,ab}
\skiplinehalf
\supit{a}
Zuse Institute Berlin (ZIB),
Takustra{\ss}e 7,
D\,--\,14\,195 Berlin,
Germany
\smallskip\\
\supit{b}
JCMwave GmbH,
Bolivarallee 22, 
D\,--\,14\,050 Berlin,
Germany
}
\authorinfo{
Corresponding author: S.~Burger\\
URL: http://www.zib.de/en/numerik/computational-nano-optics.html\\
URL: http://www.jcmwave.com\\
Email: burger@zib.de
}

\begin{document}
\maketitle

\noindent
This paper will be published in Proc.~SPIE Vol. {\bf 8255}
(2012) 82550K 
({\it Physics and Simulation of Optoelectronic Devices XX; 
Bernd Witzigmann, Marek Osinski, Fritz Henneberger, Yasuhiko Arakawa, Editors}, 
DOI: 10.1117/12.906372),
and is made available 
as an electronic preprint with permission of SPIE. 
One print or electronic copy may be made for personal use only. 
Systematic or multiple reproduction, distribution to multiple 
locations via electronic or other means, duplication of any 
material in this paper for a fee or for commercial purposes, 
or modification of the content of the paper are prohibited.

\begin{abstract}
We present a finite element method (FEM) solver for computation of optical
resonance modes in VCSELs. 
We perform a convergence study and demonstrate that high accuracies 
for 3D setups can be attained on standard computers. 
We also demonstrate simulations of thermo-optical effects in VCSELs. 
\end{abstract}


\section{Introduction}
\label{intro}
Vertical-cavity surface-emitting lasers (VCSELs) are light sources
with unique properties and potential applications of great interest~\cite{IGA00}.
Solution of Maxwell's equations for realistic 3D VCSELs is a challenging task~\cite{DEM10}. 
Since the VCSEL resonator is realized by distributed Bragg reflectors (DBR), 
the geometry inherits a pronounced multiscale structure. 
The devices are relatively large (thousands of cubic wavelengths)
and contain subwavelength DBR layers, 
very thin active zones and structured apertures.  
Further, the infinite exterior adjacent to the VCSEL containing also a layered structure has to be modeled 
to obtain realistic predictions of radiation loss and lasing threshold. 
A variety of methods has been used to compute optical VCSEL 
modes. 
These include FEM, finite difference time-domain methods (FDTD), 
modal expansion and approximative methods~\cite{BIE01,NYA07,DEM10,FA11}.
In most standard approaches the optical problem is restricted to purely 1D or to cylindrically 
symmetric structures. Nevertheless, many realistic 3D devices cannot be restricted in this way. This is the reason why reliable full 3D simulations become so important. Accuracy limitations of state-of-the-art 3D solvers, including also FEM solvers, 
have recently been discussed~\cite{DEM10,FA11}.

The finite element method is very well suited for simulation of nano-optical systems and 
devices~\cite{POM07,KAR09}. 
Its main features are the capability of exact geometric modeling due to usage of unstructured 
meshes and high accuracy at low computational cost. 
The finite element method offers great flexibility to approximate the solution: 
different mesh sizes $h$ and polynomial ansatz functions of varying degree $p$ can be combined 
to obtain high convergence rates. 
As a result, very demanding problems can be solved on standard workstations~\cite{BUR09b}.
We demonstrate that a FEM eigenmode solver with higher-order finite elements, adaptive
meshing and a rigorous implementation of transparent boundary conditions is 
a powerful method for 3D VCSEL mode computation. 

\section{Mathematical Background}
The main physical effects in a VCSEL are associated to time scales 
ranging over several orders of magnitude. 
Since the frequency of the optical modes is much higher than those of all other effects, 
a time-harmonic ansatz for the electric field $\Field E(x,y,z)$ is well-justified:
\begin{eqnarray}
  \Field E(x,y,z,t)&=e^{-i\omega t}\Field E(x,y,z),
\end{eqnarray}
where $\omega$ denotes the frequency. 
Using this ansatz in Maxwell's equations, the following second order equation for the electric field can be derived:
\begin{eqnarray}
  \label{eq:mwE}
\curl\Tensor{\mu}^{-1}\curl \Field{E}&=\omega^{2}\Tensor{\varepsilon}\Field{E}.
\end{eqnarray}
In this equation no exterior current or charge density sources are present.
Physically, the light field of a VCSEL is created by coupling of the electron system in the active layer
to the eigenmodes of the structure. 
In Maxwell's equations this usually enters via the complex permittivity tensor $\Tensor{\varepsilon}$
(in all relevant optical materials the magnetic permeability $\Tensor{\mu}$ is a constant). 
The resonance problem then consists of finding pairs $\left(\Field E,\omega\right)$, 
such that Maxwell's equations (\ref{eq:mwE}) on the given geometry are fulfilled. 
Furthermore, the so called radiation condition has to be satisfied which 
requires that the resonance modes are purely outward radiating.
For solving equations (\ref{eq:mwE}) we use the FEM package 
JCMsuite developed by ZIB and JCMwave~\cite{POM07}.

\begin{figure}[b]
\centering
\subfloat[]{\label{setup2d}
\includegraphics[width=0.62\textwidth]{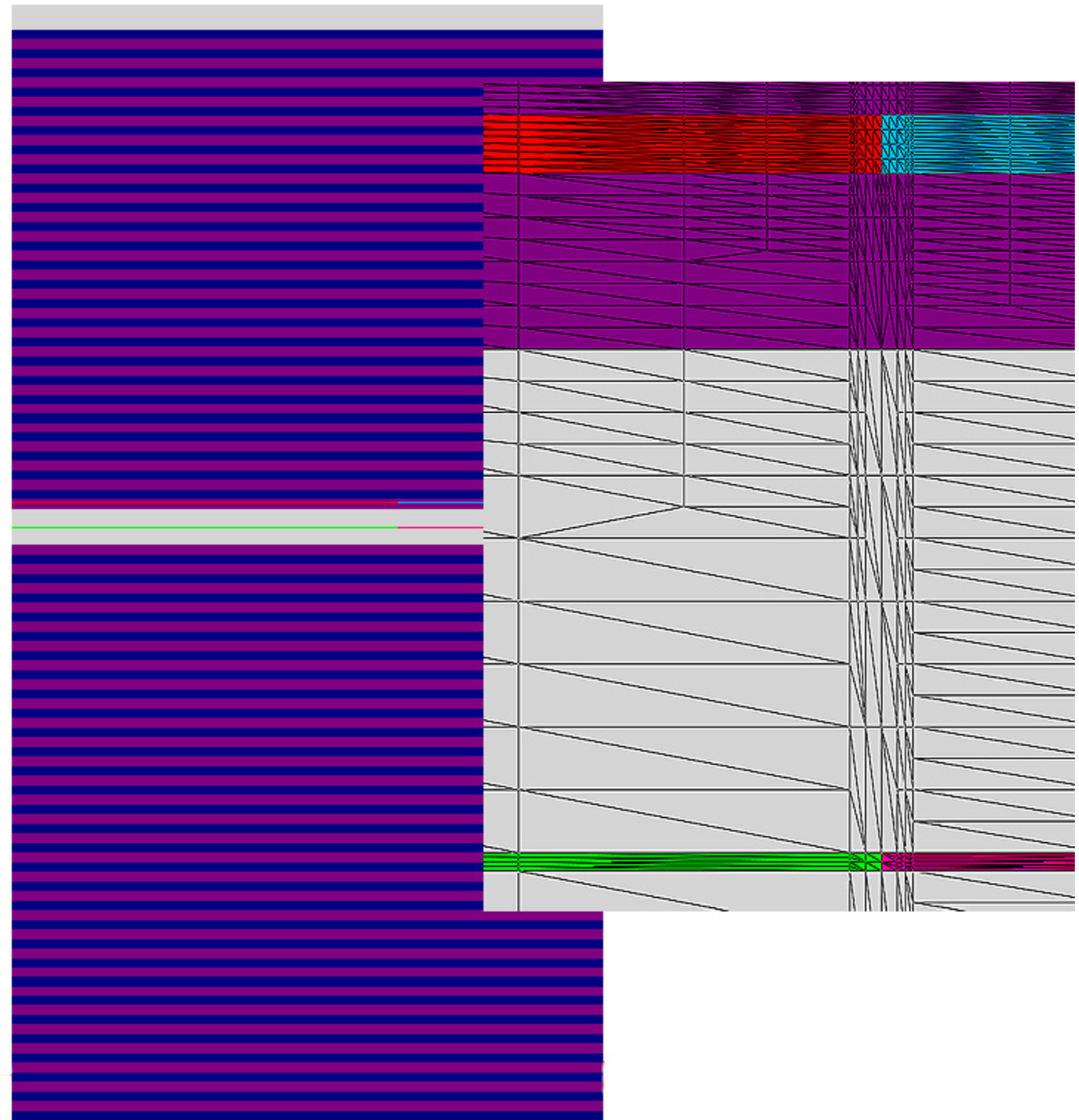}}
\hfill
\subfloat[]{\label{mode2d}
\includegraphics[width=0.33\textwidth]{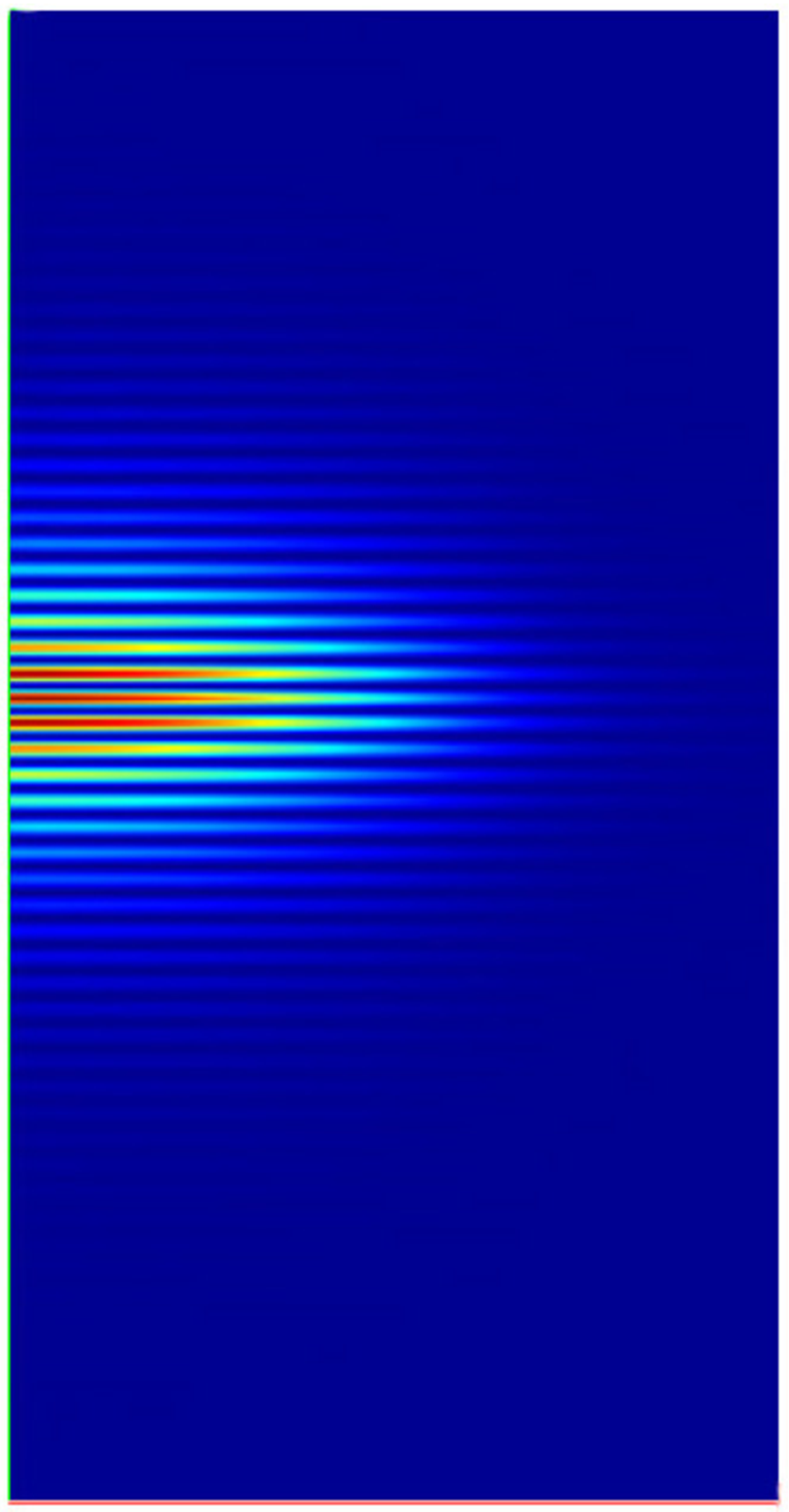}}
\caption{a) 2D cross section of cylindrically symmetric VCSEL 
(aperture diameter = \unit{6}{\micro\meter}, cf., Bienstmann et al.~\cite{BIE01}), 
the inset shows a detail of the finite element triangulation, including part of the 5\,nm thin gain region ({\it green}). 
b) Visualization of the computed electric field ($|\Field E(x,y,z)|^2$) of the fundamental mode 
($\lambda = 980.587$\,\mbox{nm}, $\gamma = -2.388\cdot 10^{-5}$).}
\label{2d}
\end{figure}

\section{VCSEL model and simulation results}

The aim of this paper is to demonstrate that very low numerical errors can be reached in full 3D VCSEL simulation.
In order to quantify the error of a numerical solution we compare it to a reference solution.
Because for the full 3D VCSEL problem no accurate quantitative results are 
available as independent reference, we use a cylindrically symmetric VCSEL setup in this convergence study.
A very accurate solution to this problem can be obtained using a 2D solver 
in cylindrical coordinates~\cite{KAR09}. 
Restriction of 3D resonance mode computations to a 2D cross section due 
to the cylindrical symmetry leads to substantial savings in computational time and memory requirements. 
The 3D, rotationally symmetric solution can also be compared to results from the literature~\cite{BIE01}.
With the reference solution at hand we perform simulation of the same physical 
setup using a full 3D FEM model. Numerical accuracy of the 3D results from this 
model is then obtained from comparison to the reference solution. 

In a further sub-section, we demonstrate 3D simulation of the thermal lensing effect in a VCSEL. 

\begin{figure}[t]
\centering
\psfrag{Rel err (wavelength)}{$\Delta\lambda$}
\psfrag{N (unknowns)}{$N$}
\psfrag{Rel err (damping)}{$\Delta\gamma$}
\subfloat{
\qquad
\includegraphics[width=0.38\textwidth]{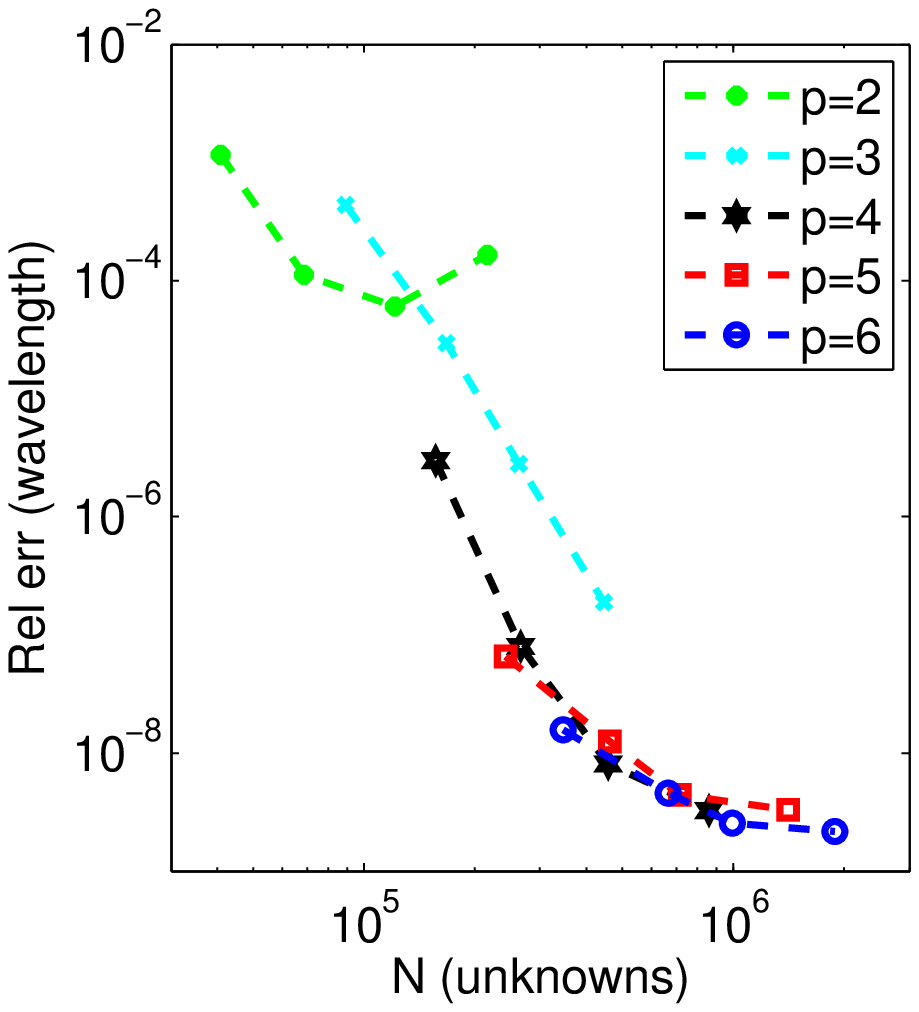}}
\hfill
\subfloat{
\includegraphics[width=0.38\textwidth]{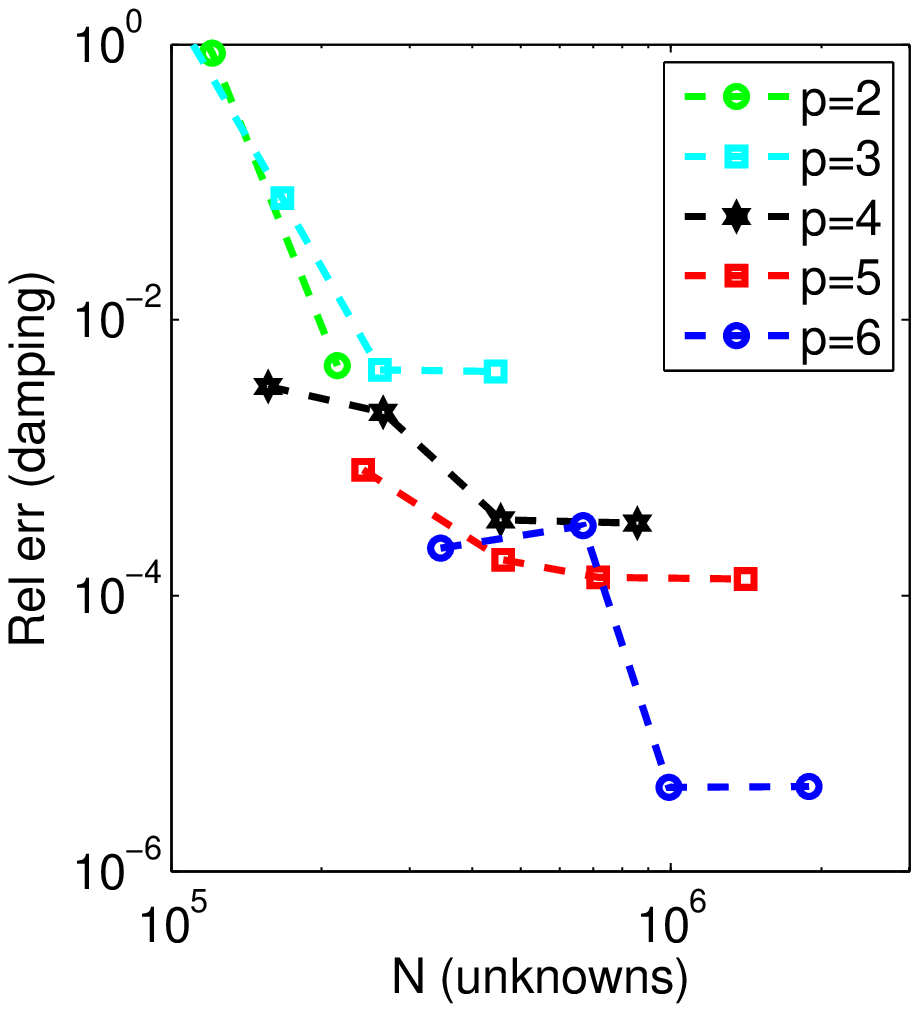}
\qquad}
\caption{Convergence graphs (reference solution): relative errors of the resonance 
wavelength $\lambda$ and the damping factor $\gamma$ in dependence on number of 
unknowns $N$ for different polynomial orders $p$ of the FEM approximation.}
\label{conv-hp}
\end{figure}

\begin{figure}[b]
\centering
\includegraphics[width=0.25\textwidth]{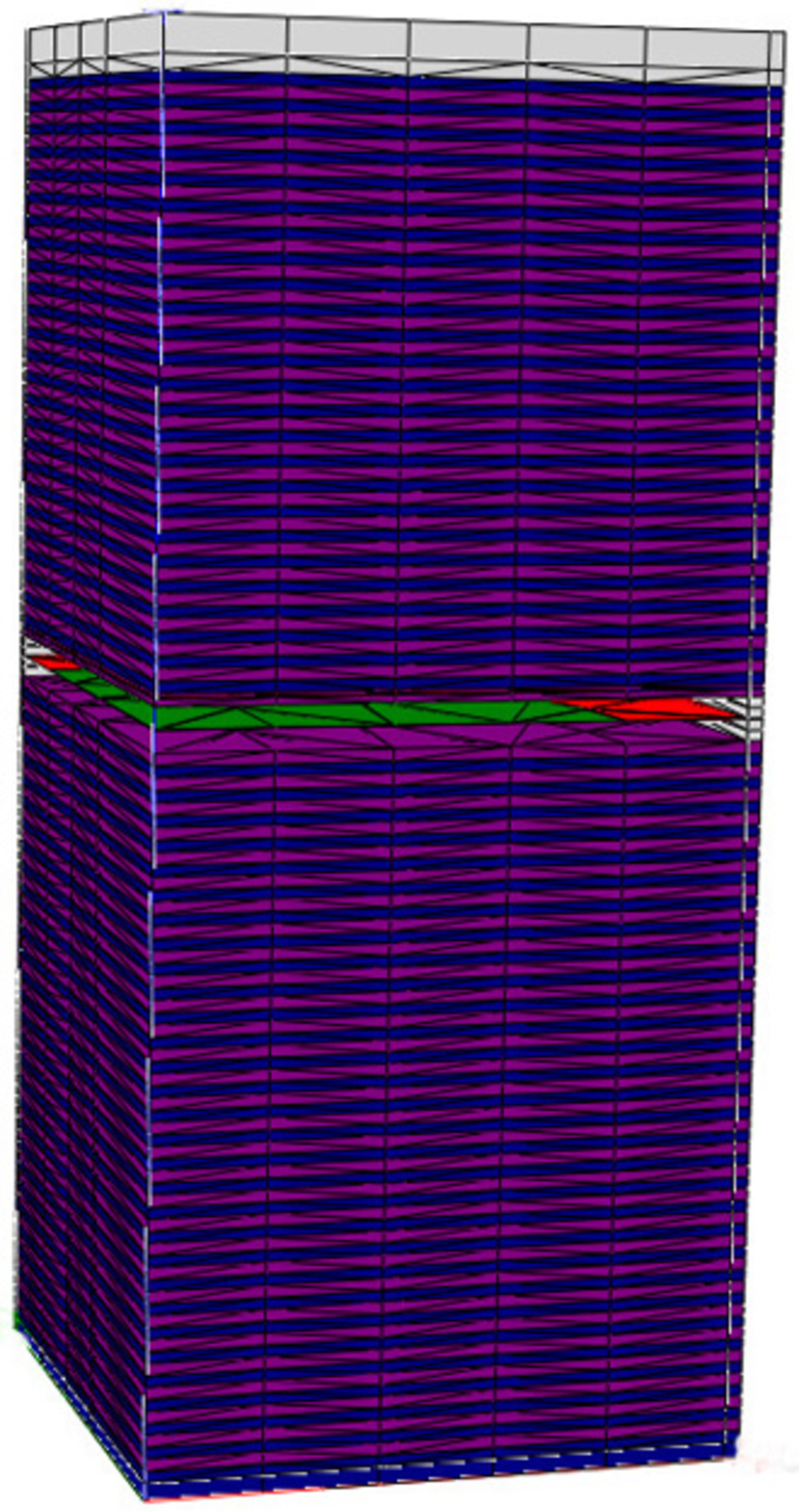}
\qquad\qquad
\includegraphics[width=0.25\textwidth]{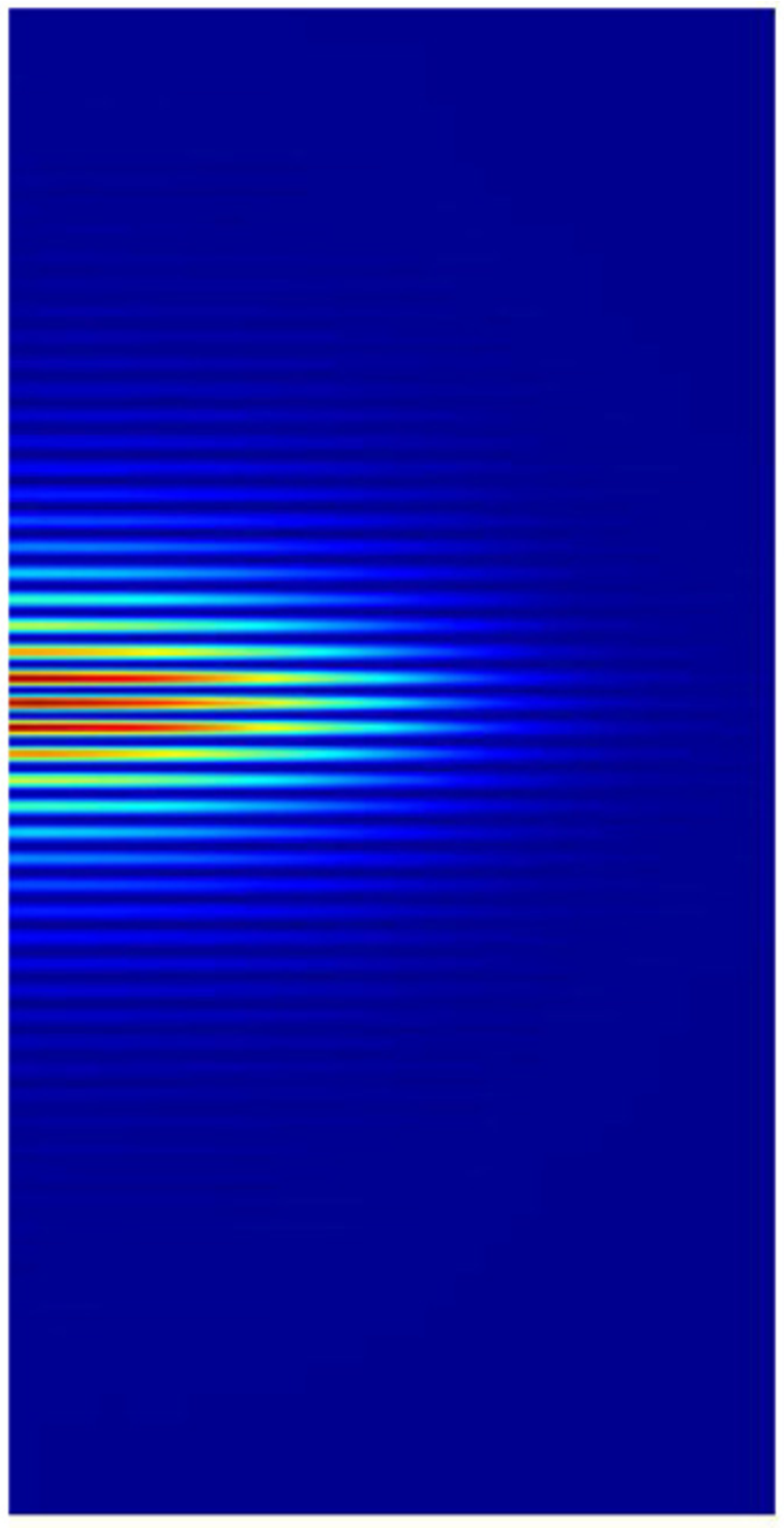}
\caption{3D layout of the cylindrically symmetric VCSEL with finite element 
triangulation ({\it left}) and visualization of the fundamental resonance mode 
($\lambda = 980.584$\,\mbox{nm}, $\gamma = -2.375\cdot 10^{-5}$, {\it right}).}
\label{3d}
\end{figure}

\subsection{2D reference solution}
As physical model we choose a VCSEL setup as described by Bienstman et al.~\cite{BIE01}. The laser consists of two DBR mirrors with alternating AlGaAs-GaAs layers. The InGaAs quantum well layer (gain material) is embedded in the central GaAs cavity region. An AlOx aperture is placed in the upper region of the lowest AlGaAs layer of the top mirror. The whole structure is situated on a GaAs substrate. Figure~\ref{setup2d} shows a 2D cross section through the cylindrically symmetric setup. Note that the layered structure extends infinitely in radial direction and modal confinement is reached through the finite AlOx aperture and the finite active region. 

We use a 2D FEM solver in cylindrical coordinates to obtain the near field solution $E$ and the complex eigenfrequency $\omega$. From $\omega$ we derive the resonance wavelength $\lambda=c\cdot {2\pi}/{\Re(\omega)}$, where $c$ denotes the speed of light, and the damping factor $\gamma={\Im(\omega)}/{\Re(\omega)}$, which quantifies the gain/loss of the cavity mode. Figure~\ref{mode2d} shows the electric field intensity distribution of the fundamental VCSEL mode in a pseudo-color representation.

We have computed solutions to the same physical setup using different numerical resolutions, where we have varied finite element degree $p$ and the mesh refinement. With increasing $p$ and increasing mesh refinement also the number of unknowns of the FEM problem, $N$, increases. Using the solution with highest $p$ and highest refinement ($p=6$, 4 successive adaptive grid refinements) as reference we can attribute relative numerical errors to the computed resonance wavelengths and damping factors. Figure~\ref{conv-hp} shows how these converge with increasing $N$. 
We have also checked convergence with respect to the numerical parameters 
of the perfectly matched layers (PML)~\cite{POM07}.
We observe that very high accuracies can be reached with relative errors below $10^{-8}$ for the resonance wavelength. 
Note that the relative error of the damping factor is larger due 
the much smaller total value of the imaginary part of the fundamental eigenvalue. 
However, also here we reach relative accuracies below 0.001\%.

\subsection{Full 3D computation}
\label{full3dsection}
For full 3D computation of the same VCSEL setup as before we discretize the 
VCSEL geometry using a 3D prismatoidal mesh. Figure~\ref{3d} (left)
shows a visualization of the mesh discretizing the geometry. 
Note that due to mirror symmetries of the device we can restrict the computational domain to 
one quarter of the full VCSEL, on the mirror faces 
boundary conditions corresponding to the symmetry of the fundamental mode are applied. 
Further, transparent boundary conditions, realized with perfectly matched layers are applied.
Figure~\ref{3d} (right) 
shows a cross section through the computed 3D field distribution ($|\Field E(x,y,z)|^2$). 
As expected the field distribution agrees perfectly with the field distribution 
of the reference solution and the resonance wavelengths agree to high precision.

\begin{figure}[t]
\centering
\psfrag{Relative err}{$\Delta\lambda\vert\Delta\gamma$}
\psfrag{N (unknowns)}{$N$}
\psfrag{d l}{$\Delta\lambda$}
\psfrag{d g}{$\Delta\gamma$}
\psfrag{gamma}{$\gamma$}
\psfrag{ni}{$n_{i}$}
\subfloat[]{\label{conv3d}
\qquad
\includegraphics[width=0.35\textwidth]{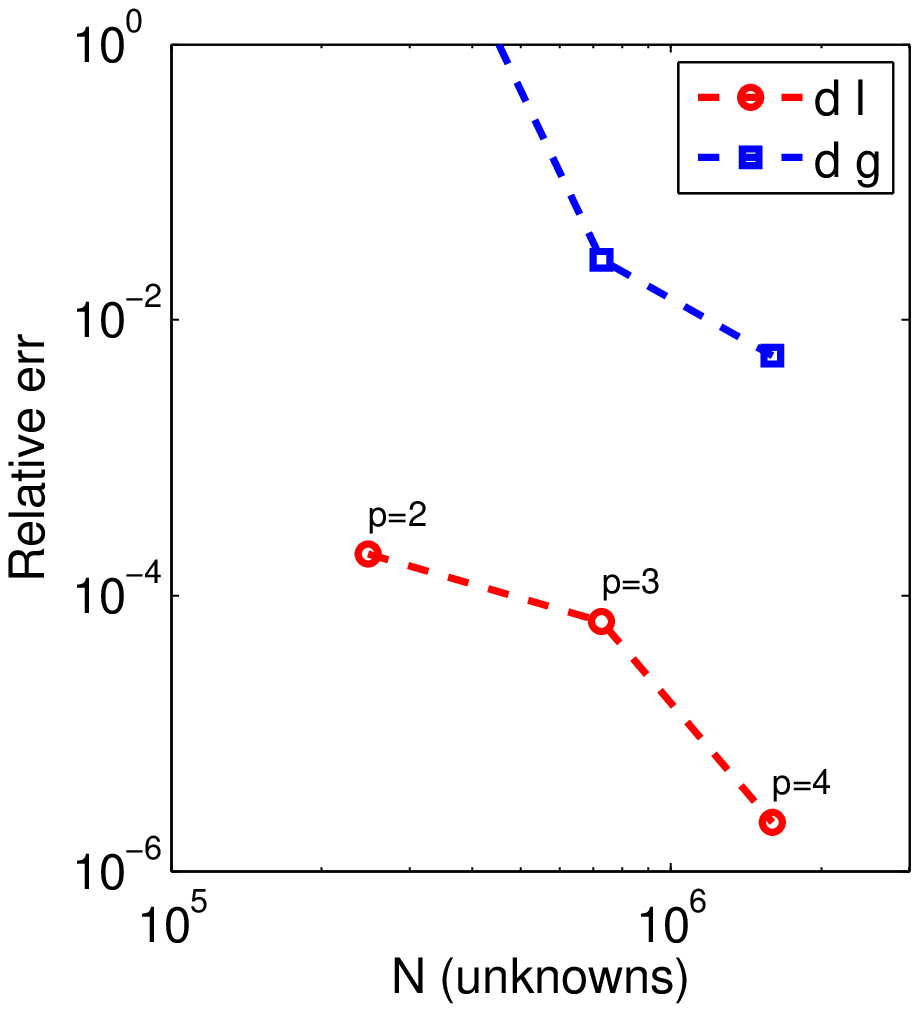}}
\hfill
\subfloat[]{\label{threshold} 
 \includegraphics[width=0.35\textwidth]{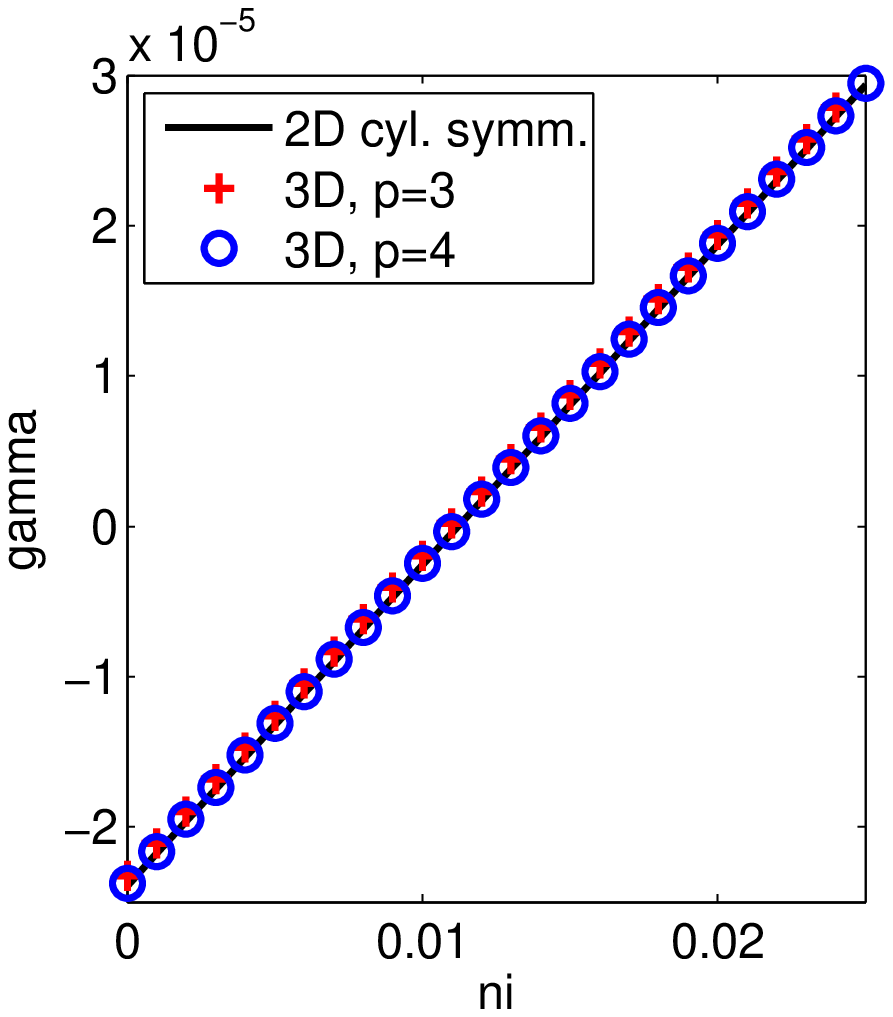}
\qquad}
\caption{a) Relative error of the resonance wavelength $\lambda$ and the damping factor $\gamma$ for 3D simulation versus cylindrical 2D solution in dependence on number of unknowns $N$ for different polynomial orders $p$ of the FEM approximation. b) Threshold gain for 2D and 3D simulations for different polynomial orders $p$.}
\label{conv3Df}
\end{figure}

Again we perform a convergence study where we vary finite element degree $p$. 
The relative errors are defined as deviations from the 2D cylindrically symmetric reference solution (see above), 
normalized to the reference solution result. 
Figure \ref{conv3Df} shows that we reach an accuracy of the resonance wavelength well below $10^{-5}$ and an accuracy of the damping factor below $10^{-2}$. Computational times for the full 3D simulations are few minutes on a standard workstation (3\,min ($p=2$), 6\,min ($p=3$), resp. 30\,min ($p=4$)).

\begin{figure}[t]
\centering
\subfloat{
\includegraphics[width=0.4\textwidth]{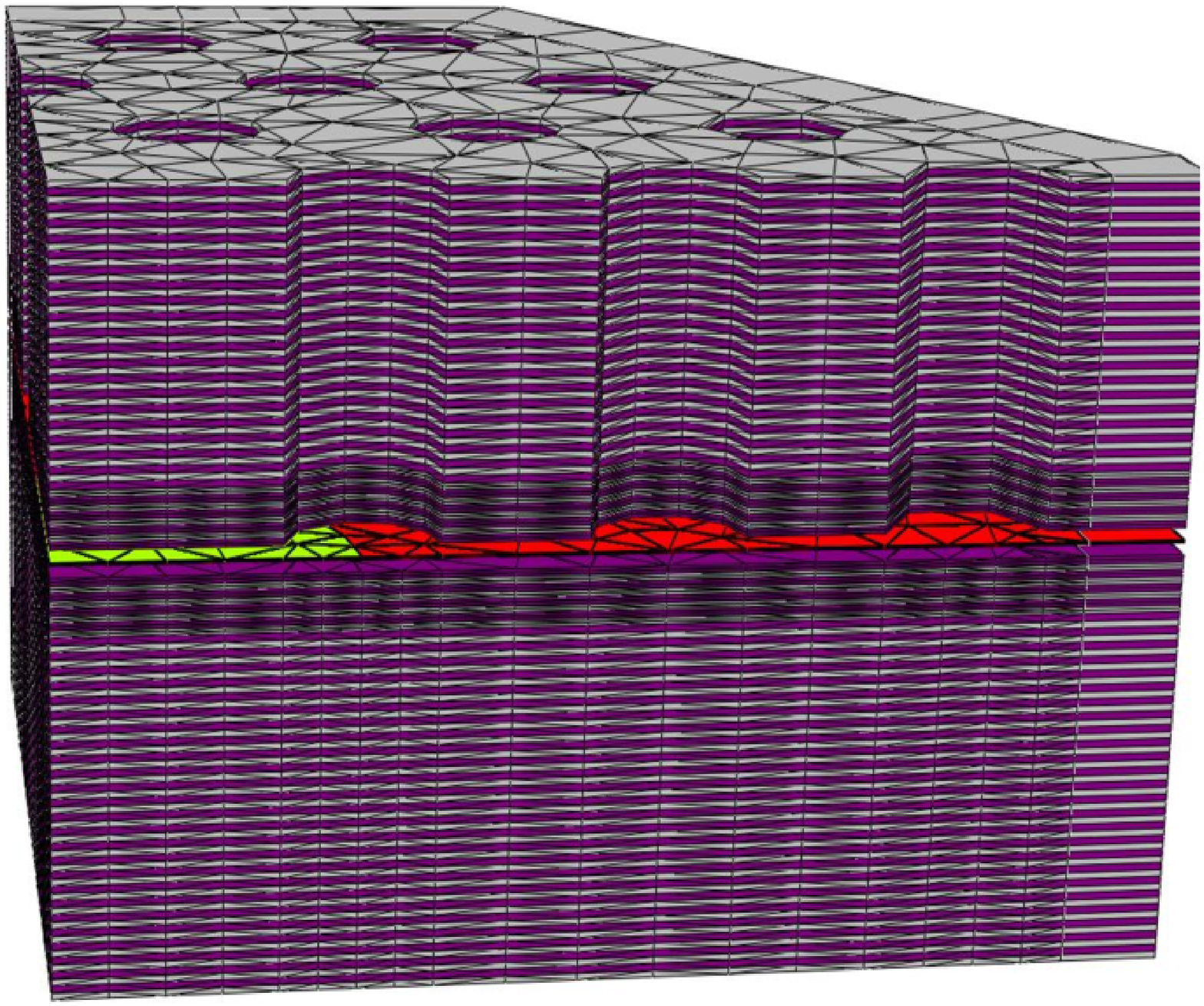}}
\hfill
\subfloat{
\includegraphics[width=0.58\textwidth]{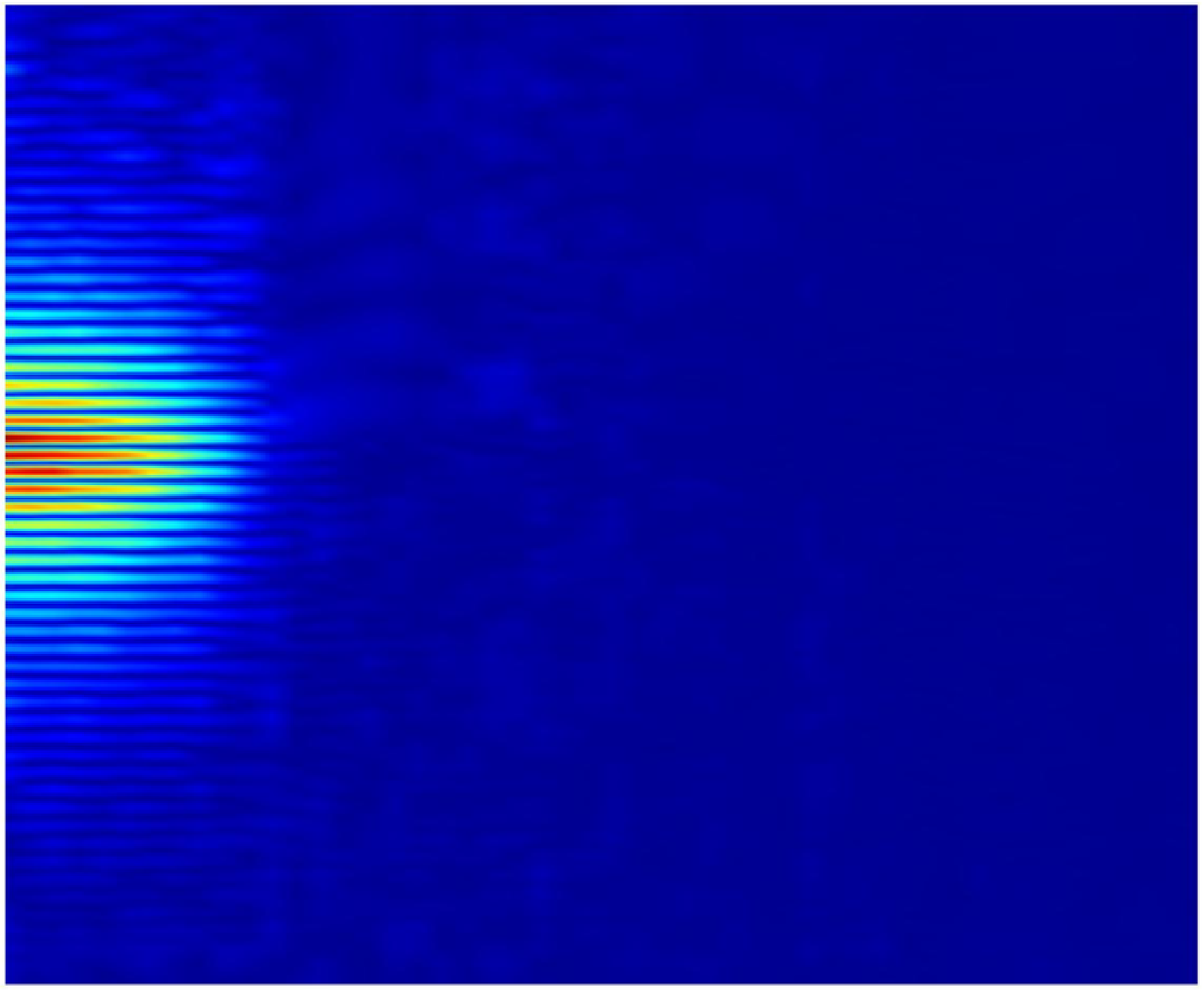}}
\caption{ 3D layout of a photonic-crystal VCSEL with finite element triangulation 
({\it left}) and visualization of its fundamental resonance mode ($|E_x|$, 
$\lambda = 978.92\,$\mbox{nm}, $\gamma=-7.63\cdot10^{-5}$, {\it right}).}
\label{photCrys}
\end{figure}

In the previous we have simulated a "cold cavity'' VCSEL, i.e., without gain in the active zone. Next we demonstrate the accuracy of laser threshold computation. For this we perform computations with increasing gain in the active zone (imaginary part of the complex refraction index of the quantum well layer, $n_i$).
Figure~\ref{threshold} shows the expected linear dependence of the damping factor on material gain $n_i$. 
We observe that threshold is reached at a refractive index $n_i\approx 0.011$. 
This value is obtained from 3D simulation at both investigated accuracy levels, 
as well as from the 2D cylindrically symmetric reference simulation. 
The agreement corresponds to the accuracy of the damping factor better than 1\% 
relative deviation as displayed in Fig.~\ref{conv3Df}.

Now, that we have characterized the accuracy level which can be reached by the 
3D FEM solver for typical VCSEL setups we turn to a ``real'', i.e., non-rotationally-symmetric
VCSEL geometry. 
Figure~\ref{photCrys} shows a visualization of the mesh and the 
fundamental mode of a photonic-crystal VCSEL in a setup as described in~\cite{DEM10}. 
We have used simulation accuracy settings as in the previous results. 
We therefore expect that the numerical accuracy of the wavelength is of the 
order of $10^{-5}$ and the numerical accuracy of the damping factor is 
of the order of $10^{-2}$, c.f., Fig.~\ref{conv3Df}.
This is a significantly higher accuracy than reached in the benchmark of Dems et al.~\cite{DEM10}. 
A main difference in our implementation, compared to the FEM implementation in Dems et al.~\cite{DEM10}, 
is the usage of higher-order elements, and of adaptive transparent 
boundary conditions~\cite{POM07}. 

\subsection{Thermo-optical simulations}
\begin{figure}
\centering
\subfloat{
\quad
\includegraphics[height=0.36\textheight]{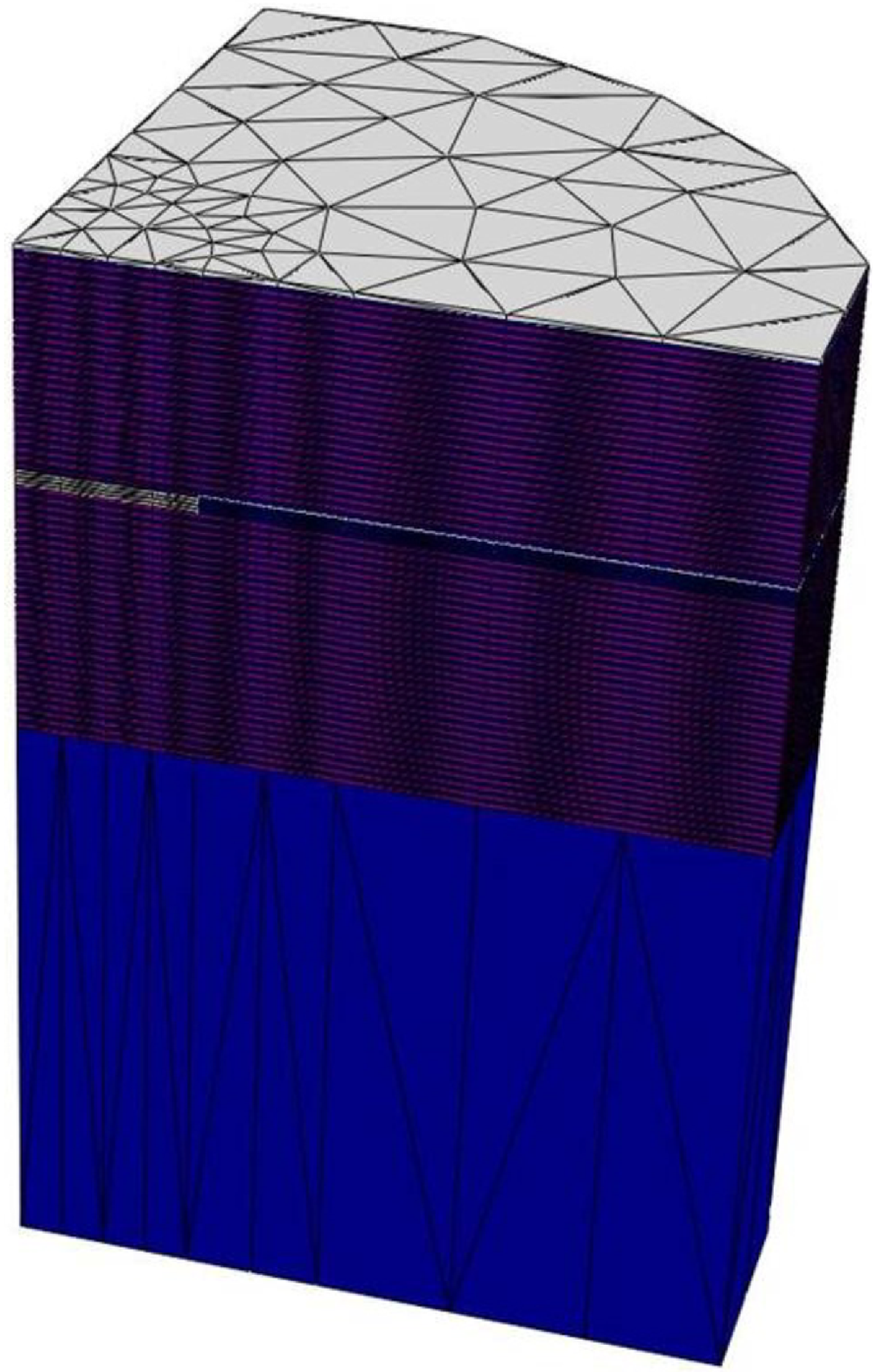}}
\hfill
\subfloat{
\includegraphics[width=0.56\textwidth]{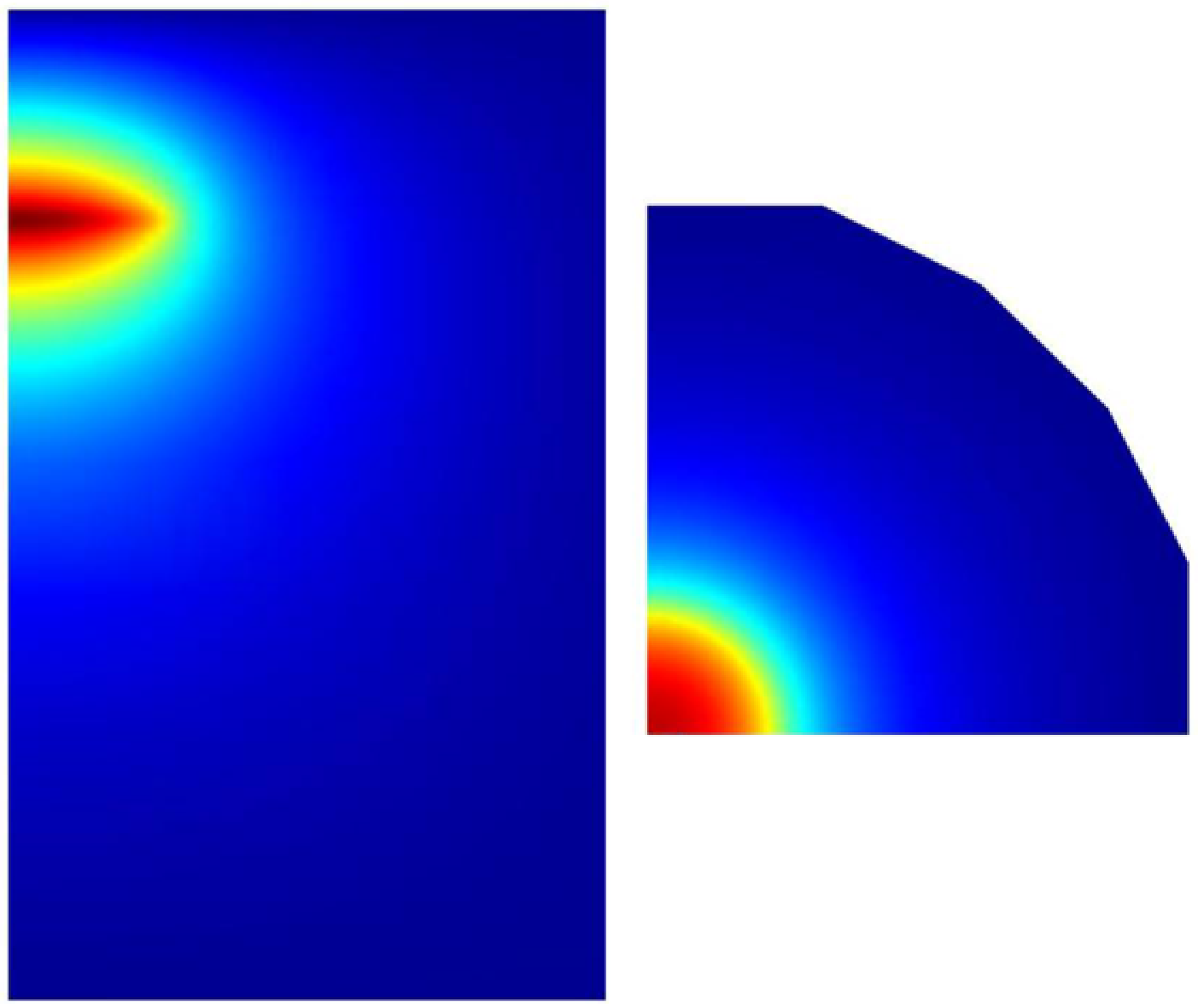}
\quad}
\caption{ 3D layout of a VCSEL for thermo-optical simulation with finite element triangulation ({\it left}).
Visualization of the computed temperature profile in an $xz$-plane ({\it center}) and in an $xy$-plane ({\it right}) 
for $\Delta T = T_{max}-T_0=50.5$K in jet colormap ($T_{max}=350.5$K (dark red), 
$T_{min}=300$K (dark blue), $T_{max}$ is the maximum temperature in the computational domain).}
\label{tempProf}
\end{figure}

\begin{figure}
\centering
  \includegraphics[height=0.3\textheight,width=0.5\textwidth]{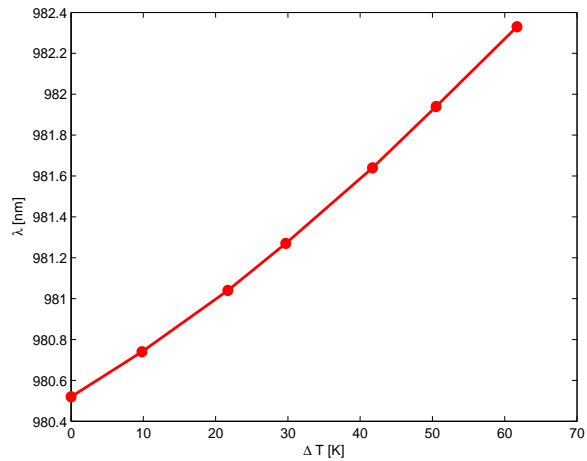}
\caption{Resonance wavelength of the optical mode, $\lambda$, 
in dependence on the maximum temperature difference $\Delta T=T_{max}-T_0$ in the VCSEL.}
\label{tempLambda}     
\end{figure}

Limitations of the output profile quality of VCSELs can be due to unconverted pump power in the active region which 
leads to a local heating and through thermo-optical coupling to inferior laser mode properties. 
In this subsection we present 3D thermo-optical simulations based on the stationary heat conduction equation
\[
-\nabla\cdot k\nabla T(x,y,z)=q(x,y,z),
\]
where $k$ denotes the thermal conductivity, $T$ denotes the temperature and $q$ is the thermal source density. 
We couple this equation to the optical simulation via the thermo-optical correction of the temperature 
dependent relative permittivity $\epsilon(T)=\epsilon(T_0)+C(T-T_0)$, 
where $C$ is the thermo-optical coefficient and $T_0$ is a reference temperature. 

\begin{figure}[t]
\centering
\subfloat[]{\label{E_vert}
\includegraphics[height=0.3\textheight,width=0.45\textwidth]{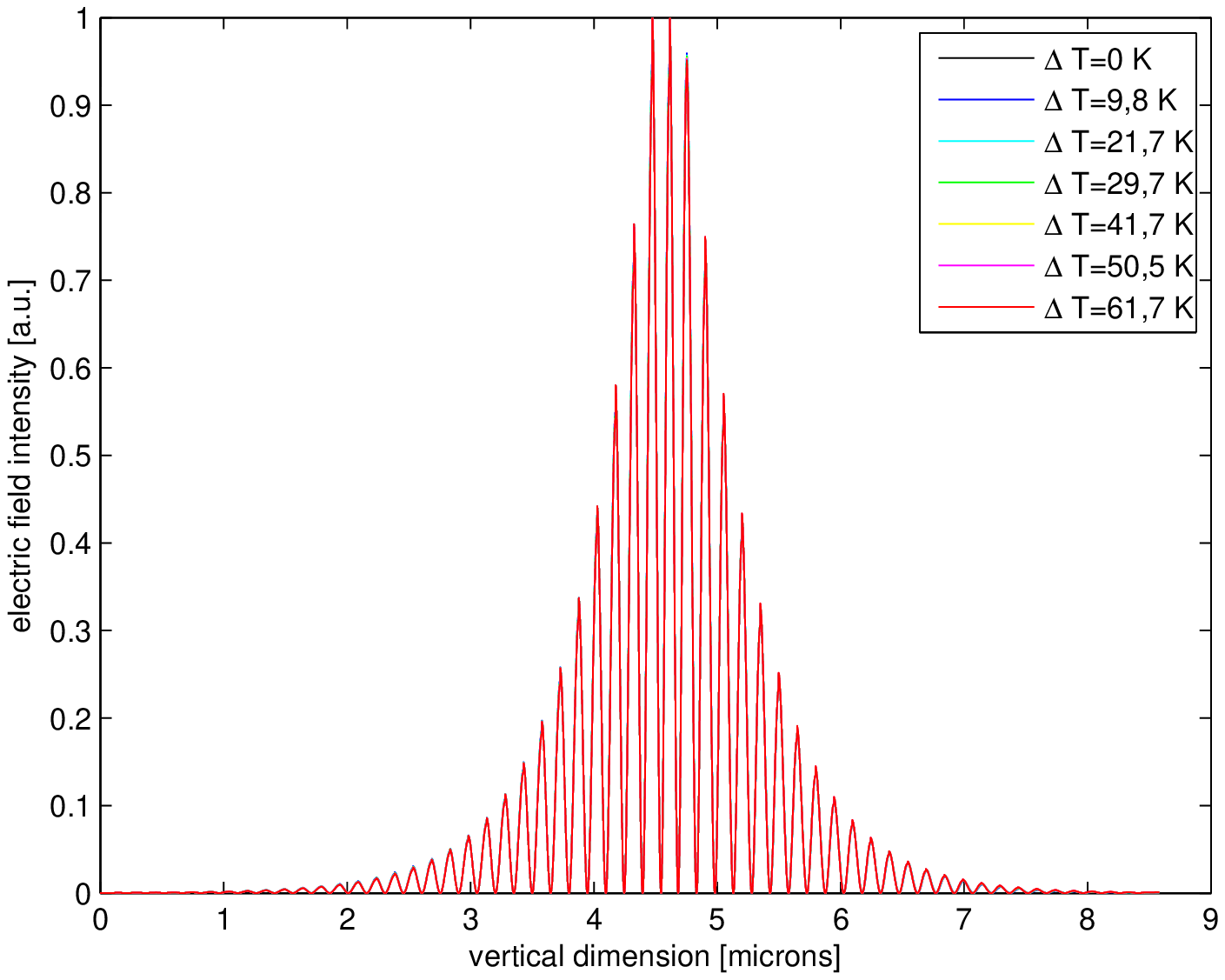}}
\hfill
\subfloat[]{\label{E_hor} 
 \includegraphics[height=0.3\textheight,width=0.45\textwidth]{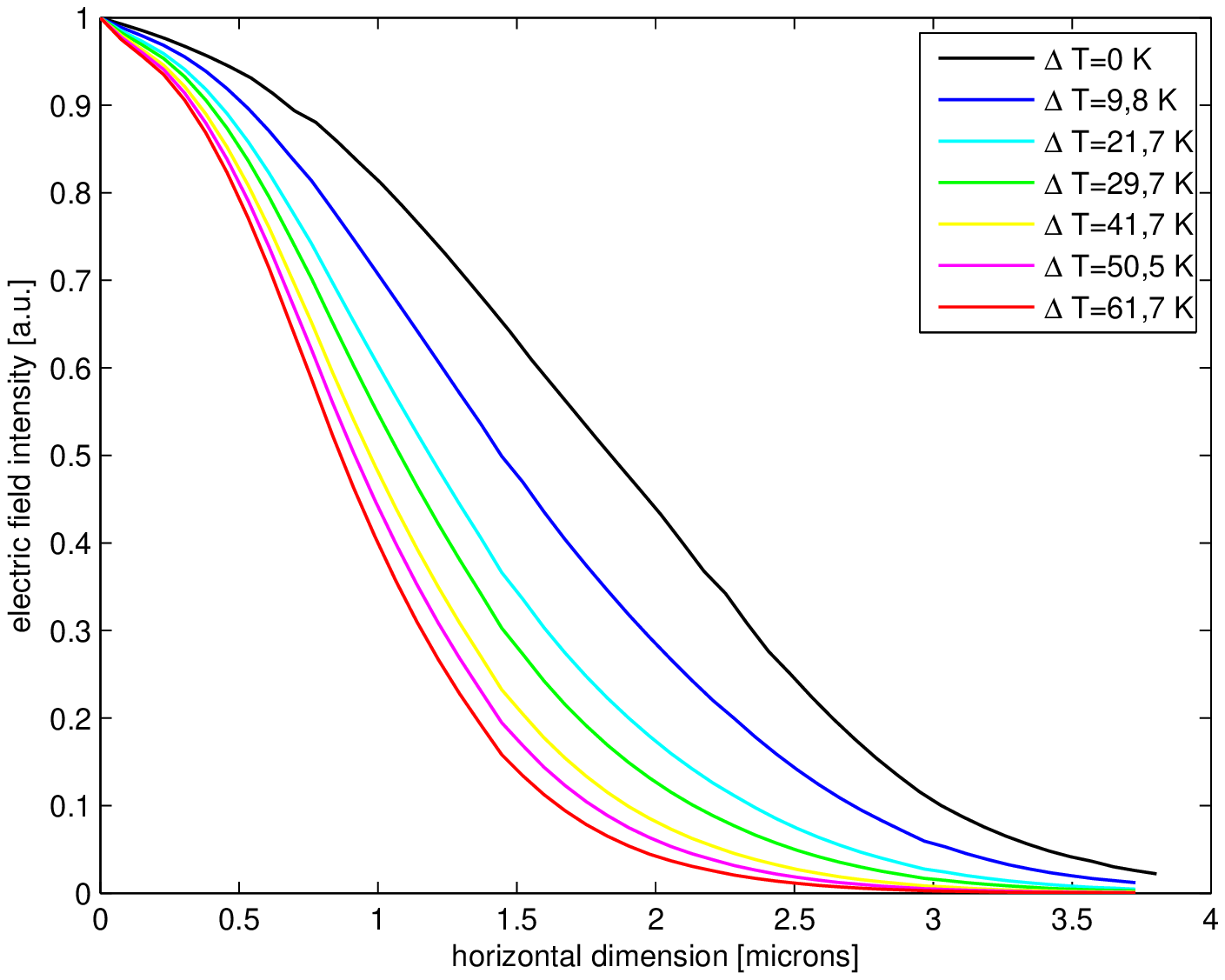}}
\caption{a) Electric field intensity along the vertical dimension. 
b) Electric field intensity along the horizontal dimension.}
\label{E}
\end{figure}

We investigate the same VCSEL geometry and material parameters 
as in the previous Section~\ref{full3dsection}. 
For the thermal simulation 
we need to expand the computational domain (GaAs substrate) in order to set suitable boundary conditions, 
i.e., to fix the boundary temperature of the computational domain, $T_0$, at $300$K. 
We define a heat source in the oxide aperture and in the cavity by setting the thermal source density $q$ 
in the according regions and compute the temperature distribution within the VCSEL by solving the stationary
heat equation. Fig.~\ref{tempProf} shows visualizations of the discretized geometry and of a computed 
temperature distribution. 
We then use this temperature distribution to define a locally varying permittivity correction throughout the 
VCSEL geometry, and compute the optical cavity modes using the Maxwell solver as in Section~\ref{full3dsection}. 

We have performed simulations for various heat source powers: 
Figure~\ref{tempLambda} shows the almost linear dependence of the resonance 
wavelength $\lambda$
on the maximum temperature difference in the device. 
Increasing the maximum temperature in the device through increasing heat source power 
causes a refractive index change of the DBR structure 
and a shift of its reflective wavelength. 
Thus the VCSEL resonance wavelength $\lambda$ is shifted.

In post-processes, from the near field distribution we exported 1D electric field intensity distributions 
($|\Field E|^2$) in both, vertical and horizontal directions, through the cavity center. 
The electric field along the vertical dimension is nearly independent on temperature variations, as expected, 
see Fig.~\ref{E_vert}. 
In the horizontal direction we observe the constriction of the electric field intensity with 
increasing temperature, see Fig.~\ref{E_hor}. 
Such behaviour is known as thermal lensing effect and a limitation for high-power semiconductor lasers.
In specific applications, VCSELs should be designed for low thermal lensing~\cite{WEN11}.

\section{Conclusion}
We have presented FEM results for resonance mode computation in VCSELs. 
Our results for a cylindrically symmetric device demonstrate fast 
convergence and high accuracy of the method. 
The resonance wavelength and the corresponding damping factor can be computed 
very accurately with relative errors below $10^{-8}$, respectively $10^{-5}$. 
The reached accuracies are several orders of magnitude higher than the differences between the 
results from different models presented in Ref.~\cite{BIE01}. 

Comparison of full 3D simulation results to the 2D solution and a detailed 
convergence analysis confirm the high accuracy of the method also for full 
3D problems with relative errors of the wavelength and the corresponding 
damping factor below $10^{-5}$, respectively $10^{-2}$. 
To our knowledge results on numerical accuracy of 3D VCSEL simulations on  
this level have not previously been published (see, e.g.,~\cite{BIE01,NYA07,DEM10}). 
We have also presented thermo-optical simulations for 3D VCSELs. 
In future research we plan to investigate more complex 3D device geometries 
like metal cavity surface-emitting microlasers. 

\section*{Acknowledgments}
We acknowledge funding by the {\it Deutsche
Forschungsgemeinschaft} (DFG, German Research Foundation) within 
SFB\,787 {\it Halbleiter Nanophotonik}. 

\bibliography{bibliot}
\bibliographystyle{spiebib}  

\end{document}